\documentclass[]{aa}
\usepackage{array}
\usepackage{graphicx}
\usepackage{float}
\usepackage{fixltx2e}

\usepackage{txfonts}
\begin{document}

   \title{Non-interacting coronal mass ejections and solar energetic particles near the quadrature configuration of Solar TErrestrial RElations Observatory}


   \author{Anitha Ravishankar
          \and
          Grzegorz Micha$\l$ek
          }

 \institute{Astronomical Observatory of Jagiellonian University, Krakow, Poland\\
 \email{anitha@oa.uj.edu.pl}\\
 \email{grzegorz.michalek@uj.edu.pl}
 }



  \abstract
   {We present our results on the correlation of non-interacting coronal mass ejections (CMEs) and solar energetic particles (SEPs). A statistical analysis was conducted on 25 SEP events and the associated CME and flare during the ascending phase of solar cycle 24, i.e., 2009-2013, which marks the quadrature configuration of \emph{Solar TErrestrial RElations Observatory} (STEREO). The complete kinematics of CMEs is well studied near this configuration of STEREO. In addition, we have made comparison studies of STEREO and \emph{SOlar and Heliospheric Observatory} (SOHO) results. It is well known that the CME speeds and SEP intensities are closely correlated. We further examine this correlation by employing instantaneous speeds (maximum speed and the CME speed and Mach number at SEP peak flux) to check whether they are a better indicator of SEP fluxes than the average speed. Our preliminary results show a better correlation by this approach. In addition, the correlations show that the fluxes of protons in energy channel >10 MeV are accelerated by shock waves generated by fast CMEs, whereas the particles of >50 MeV and >100 MeV energy bands are mostly accelerated by the same shock waves but partly by the associated flares. In contrast, the X-ray flux of solar flares and SEP peak flux show a poor correlation.}

\maketitle



%

\section{Introduction}

Coronal mass ejections (CMEs) are intensive outbursts of plasma and magnetic field that play a key role in causing strong geomagnetic storms on the Earth (e.g., \citealt{gosling1991}; \citealt{webb}; \citealt{St}; \citealt{cane2000}; \citealt{gopalswamy2001b}, \citealt{2002a}, \citealt{gopalswamy2002}; \citealt{srivastava}; \citealt{kim}; \citealt{moon} ; \citealt{manoharan2004}; \citealt{manoharan2006}, \citealt{2010}; \citealt{manoharan2011}; \citealt{shanmugaraju}). Space weather is mainly controlled by the strong magnetic storms and particle storms that are caused by the enhanced fluxes of protons and ions. These accelerated plasmas are known as {solar energetic particles} (SEPs). The generation of SEPs is mainly due to two phenomenon, impulsive SEP events caused by magnetic reconnection manifested as solar flares (\citealt{cane1986}) and gradual SEP events accelerated by strong shocks associated with CMEs (e.g., \citealt{cane1987}; \citealt{reames1999}; \citealt{kahler2001};  \citealt{gosling1993}).

The large, gradual, and long-lived SEP events are of particular interest as the arrival of   associated interplanetary CMEs and shocks and the enhanced SEP fluxes can cause severe damage to satellites in space and technology on the ground. SEP intensities and  CME speeds are well correlated (\citealt{kahler2001}, \citealt{Pande}). As explained by \citealt{reames2013}, the CME-driven shock scatters the ions back and forth by the resonant Alfv\'{e}n waves amplified by the accelerated protons themselves as they stream away. A good indicator of SEPs accelerated by  coronal and interplanetary shocks are type II (metric) bursts (\citealt{cliver}). In addition, the particles existing in the interplanetary medium
 that are  produced by the preceding CME provide seed particles for the primary CME (\citealt{gopalswamy2001a}, \citealt{2002a}, \citealt{gopalswamy2002b}).  This scenario is crucial for space weather forecasts since the acceleration of already existing seed particles would drive the particles farther away from the Sun. In the case of halo-CMEs this proves to be an important problem, one that helps us understand the acceleration mechanism of SEPs for accurate predictions.

The scientific literature provides several results of good correlation studies between CME speed and the peak flux of the associated SEPs; for example, \citealt{kahler2001} presented a correlation of 0.616 and 0.718 at 2 MeV and 20 MeV energy band of SEP peak flux, respectively; \citealt{Pande} obtained a 0.60 correlation between halo CMEs and the associated SEP peak flux using the \emph{SOlar and Heliospheric Observatory} (SOHO) and \emph{Geostationary Operational Environmental Satellite} (GOES) >10 MeV energy bands for SEPs. It is worth  noting that these considerations were based on average velocities of CMEs. Nevertheless, we cannot rule out the influence of other parameters that drive SEPs, such as the preexisting SEPs in the ambient medium (\citealt{kahler2001}; \citealt{gopalswamy2004}), preceding CMEs (\citealt{kahler2005}), and CME-CME interaction (\citealt{gopalswamy2012}).

Recently there have been several studies that can be used to increase our ability to forecast SEP events. \citealt{Richardson14} have analyzed properties of more than 200 individual $>$25 MeV solar proton events that occurred during the period October 2006 – December 2013, using multiple spacecraft (\emph{Solar TErrestrial RElations Observatory} (STEREO)-A, -B, and SOHO). Among other parameters, they developed a formula for predicting the proton intensity at 14 – 24 MeV based on the CME speed and solar event location using the three spacecraft observations. This approach allows us to predict the intensity of the largest most extended events reasonably well, but it fails for a large population of weaker-than-expected events. \citet{Richardson15} used the same population of SEP events to compare their correlations with kinematic parameters (speed and width) of CMEs included in the catalogs \emph{Coordinated Data Analysis Workshops} (CDAW), \emph{Computer Aided CME Tracking} (CACTus), \emph{Solar Eruptive Event Detection System} (SEEDS), and \emph{CORonal IMage Processing} (CORIMP). The most common correlation between CME speed and proton event intensity has similar values for most catalogs. It should be noted that this convergence is determined by a few large particle events associated with fast CMEs and small events associated with slow CMEs. Intermediate particle events are more scattered when speeds from different catalogs are used. In addition, they also demonstrate that quadrature spacecraft CME speeds do not improve the correlation coefficient.

It is clear that the peak intensity of an SEP event  cannot be determined by the CME speed alone. New analyses of white-light CME images enable us to improve calculations of the CME masses, and hence their kinetic energies (\citealt{Vourlidas10}). This allowed the determination of the relationships between properties of SEPs and dynamic parameters of the CME. Recently, \citealt{kahler13} used two kinetic energies of CMEs, based on frontal and center-of-mass speeds, to predict the peak intensities and other parameters of western hemisphere 20 MeV SEP events. Those correlations proved to be higher with kinetic energy based on frontal speed than those based on center-of-mass speeds. We can assume that the body of the CME is less significant than the CME front in SEP production.
Very recently, \citealt{Xie19} significantly improved previous considerations in two aspects. Using a three-dimensional CME reconstruction method and combined STEREO, SOHO, and \emph{Solar Dynamics Observatory} (SDO) {white light} (WL) and {extreme ultraviolet} (EUV) observations as constraints, they were able to determine the true radial speed of the shock and CME angular widths.
This new technique allowed them to improve the correlation coefficient between logarithmic peak intensity and the true kinetic energy for 19 – 30 MeV protons (up to 0.9) and for 62 – 105 keV electrons (up to 0.8).

The correlation of solar flare properties and SEPs is rather poor. \citealt{gopalswamy2003} have reported a weak correlation of 0.41 between the SEP and X-ray peak flux. In the statistical study of \emph{soft X-ray} (SXR) flux of the flares, \citealt{miteva} found that SEPs observed within an \emph{Interplanetary Coronal Mass Ejection} (ICME) have a better correlation than when they propagate in the ambient solar wind.  \citet{temmer} and \citet{berkebile12} have recently suggested that \emph{hard X-ray} (HXR) emission of the flare and CME acceleration peak times occur nearly simultaneously, followed by the  \citet{nipa} suggestion on the occurrence of primary acceleration of SEPs to higher energies at the flare site. Even so, the physical relationship between the flares, CMEs, and SEPs is still debatable.

In this paper we continue to determine the correlations of CME speed and the associated SEP peak flux. We use the new approach of using the instantaneous speeds of CMEs of non-interacting halo CMEs near the quadrature configuration of STEREO. The data from SOHO/\emph{Large Angle and Spectrometric Coronagraphs} (LASCO) and STEREO/\emph{Sun Earth Connection Coronal and Heliospheric Investigation} (SECCHI) (\citealt{Brueckner95}; \citealt{Howard08}) were employed to determine the kinematics of the CMEs. 
The SEP fluxes at three energy bands (>10 MeV, >50 MeV, and >100 MeV) from The GOES \emph{Energetic Particle Sensor} (EPS), part of the \emph{Space Environment Monitor} (SEM), was used to study SEP intensities. In addition, we discuss the relation of the X-ray peak flux of solar flares using GOES-14 data associated with the two phenomena.

This article is organized as follows. The data and method used for the study are described in Section 2. In Section 3 we present results of our study. Finally, the conclusions and discussions are presented in Section 4.

\section{Data and method}
In our study we used observations from a few instruments and a new approach for determining CME speeds. In the following subsection we describe the method  we used for the purpose of our study.
\subsection{Data}
In our study, the SEPs associated with non-interacting halo CMEs were taken into consideration. The CME data were obtained from the observations of instruments on board two separate spacecraft: the LASCO instrument on board SOHO and the SECCHI instrument suite on board STEREO. The widely used SOHO/LASCO catalog\footnote{cdaw.gsfc.nasa.gov/CME$\_$list}, which contains the basic attributes of CMEs, (\citealt{yashiro2004}, \citealt{gopalswamy2009}), was used in our study. When the accuracy of determination of true speeds of CMEs is discussed, it is worth noting that the coronagraphic observations of SOHO/LASCO are subject to projection effects. Due to the projection effect the speeds obtained provide an inaccurate forecast of geoeffective events originating from the disk center. The quadrature configuration of STEREO with respect to the Earth, during the ascending phase of the solar cycle 24 (2009 - 2013), offered a big advantage in the accurate determination of plane-of-sky speeds which are close to the true radial speed of halo events (\citealt{bronarska}). Manual measurements of height-time data points were performed with the data of coronagraphs, COR1, COR2, and the heliospheric imagers, HI1 and HI2, which are part of SECCHI, data available on \emph{UK Solar System Data Centre (UKSSDC)\footnote{https://www.ukssdc.ac.uk/solar/stereo/data.html}}, to determine the speed of CMEs. To obtain the most accurate height-time data points, we employed only the images from STEREO-A or -B, which showed better quality. Before making this selection for each event we checked images from  both satellites. Among the 25 events, STEREO-A showed good quality for 24 events except for the event on 31 August 2012. This event   originates from the east hemisphere and therefore STEREO-B is the spacecraft with a better “side view” of the CME.

During the ascending phase of solar cycle 24 (i.e., 2009-2013) 46 large SEP events with flux $\geq$1 pfu (1 pfu = 1 proton cm{$^{-2}$} s{$^{-1}$} sr{$^{-1}$}) in the >10 MeV energy band were recorded (list from \citealt{gopalswamy2015};  \citealt{Pande}); 21 events were excluded as they were interacting or too faint to observe by STEREO.
Interaction occurs when consecutive fast events hit previous slower events. During these interactions additional flux of particles can appear. These are complex phenomena, difficult to study using statistical methods since they require a case by case analysis.
The SEP fluxes were observed by the SEM instrument on board GOES-13 geostationary satellite recorded in the \emph{National Oceanic and Atmospheric Administration} (NOAA) database\footnote{https://satdat.ngdc.noaa.gov/sem/goes/data/avg/}. The energetic protons in the energy bands >10 MeV, >50 MeV, and >100 MeV were used for this study. We selected those events when the proton flux is larger than the average background flux. In the >10 MeV energy band, events with flux value $\geq$1 pfu  and in the >50 MeV and >100 MeV bands, events with $\geq$0.1 pfu flux value were selected for the study. The associated X-ray data were obtained from \emph{Solar X-Ray Sensor (XRS) event files of GOES-15 satellite observations on the NOAA  database\footnote{https://satdat.ngdc.noaa.gov/sem/goes/data/full/}.} The soft X-ray flux events recorded in the 1–8 Å energy band were considered for the study. The Hinode  Flare Catalogue\footnote{https://hinode.isee.nagoya-u.ac.jp/flare$\_$catalogue/} (\citealt{watanabe}) provides data of the onset time, peak time, and the class of solar flares.

Based on the above conditions of CMEs and SEPs, we obtained 25 good events, which are listed in Table 1. Figure~1 shows the heliographic locations of the active regions associated with 25 non-interacting halo CMEs considered in our study. We observe that they mostly originate from the west limb or disk center. In our sample, one CME is located on the east side of the solar disk. There were four backside flare events whose fluxes could not be determined. \citealt{gleisner} reported that the characteristics of the observed SEP fluxes are determined by the strength and spatial structure of the shock. The shock formation near the Sun is usually indicated by the onset of type II bursts. The date and time of these shocks can be obtained from Wind/WAVES  type II burst catalog\footnote{https://cdaw.gsfc.nasa.gov/CME$\_$list/radio/waves$\_$type2.html}(\citealt{bougeret}). The catalog includes DH type II bursts determined using the data of WIND/Waves and STEREO.


    \begin{figure}[h!]
\includegraphics[width=9cm,height=6cm]{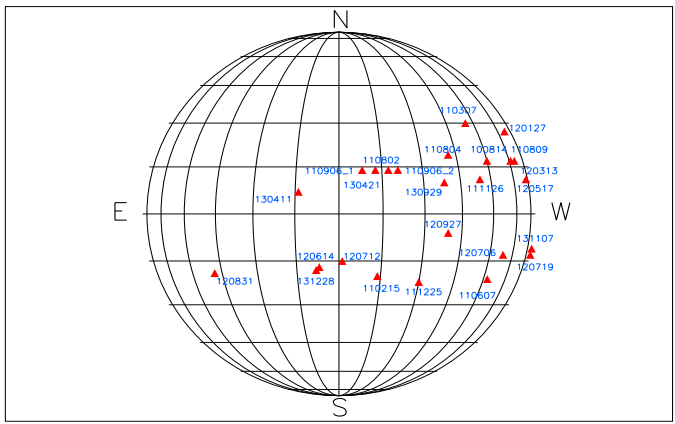}
   \caption{Heliographic locations of the active regions associated with 25 non-interacting halo CMEs.}
              \label{FigGam}%
    \end{figure}

    \begin{figure*}[h!]
   \centering
   \includegraphics[width=14cm,height=10cm]{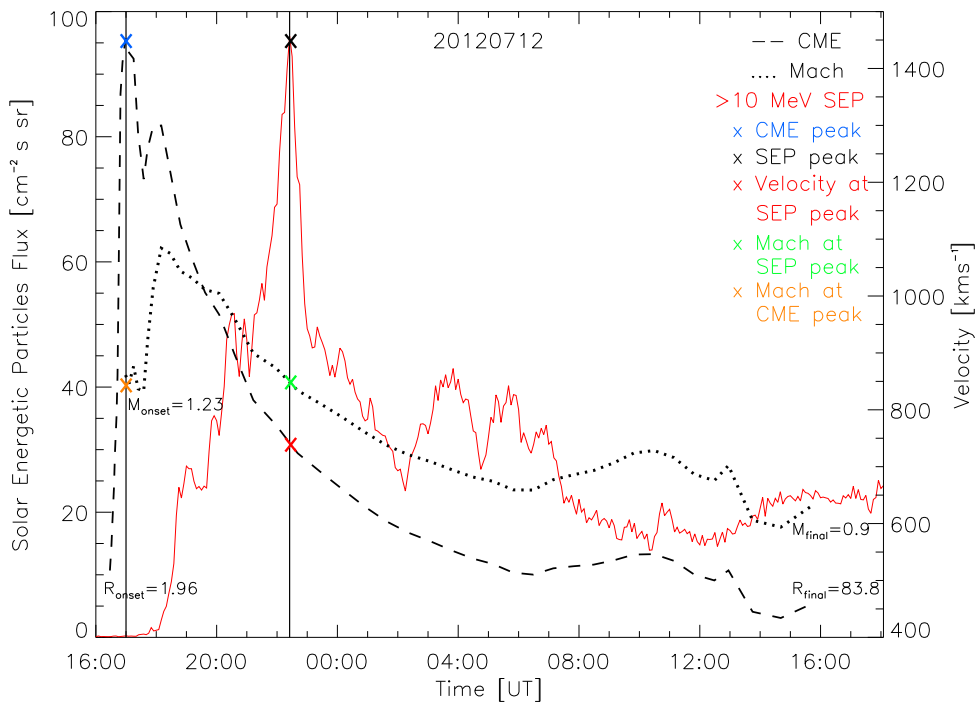}
   \caption{12 July 2012 event of a CME and the associated SEP. The plot shows the CME speed from STEREO (dashed line), Mach number (dotted line), and SEP flux in the >10 MeV energy band (red line) vs time. The Mach number has been scaled by (100) to match the plot.
   The CME peak velocity, SEP peak flux, velocity of the CME at the SEP peak flux, and Mach number at the SEP peak flux and at the CME peak velocity are shown as  crosses for reference (blue, black, red, green, and orange, respectively). The onset and final values of Mach number (M\textsubscript{onset}, M\textsubscript{final}) and distance from the Sun [ in units R\textsubscript{sun}] (R\textsubscript{onset}, R\textsubscript{final}) are shown in the figure.}
              \label{FigGam}%
    \end{figure*}

 \subsection{Method}
 In the previous studies, average speeds of CMEs were considered in the determination of correlation with the associated SEPs. Unfortunately, CMEs move at variable speeds. Hence, this technique only approximates the relationship between CME properties and fluxes of SEP.
 Therefore, in the present research we use a different approach to improve the accuracy. In the study we associate instantaneous ejection speeds with SEP fluxes.
For this purpose, for all considered CMEs, we determined velocity profiles depending on the time and distance from the Sun. The velocity profiles, for comparison, were obtained for both SOHO and STEREO observations. For SOHO observations, we used the height--time points from the SOHO/LASCO catalog, and for the STEREO observations we determined the height--time points ourselves.
 To determine the instantaneous velocity profiles we used linear fits to five height--time data points. Shifting successively these five points linear fits, we obtain the instantaneous profiles of speed in time and in distance from the Sun (\citealt{ravishankar}). Technically, two neighboring height--time points are enough to determine the  instantaneous speed, but as manual measurements are subject to unpredictable random errors, we used five successive points to obtain the most reasonable results. In addition, we applied a linear fit to the height--time data points to further minimize the impact of errors on the determined instantaneous speed. Figure~2 displays an example of the time and distance variation of CME speed for STEREO observations. Additionally, we added variation of Mach number of CME and flux of SEP for three energy channels. Instantaneous velocities that are also used in our study are clearly shown in figure. The peak speed, the peak SEP flux, the speed of the CME at the peak SEP flux, Mach number at the peak CME speed, and that at the peak SEP flux are shown as  blue, black, red, orange, and green crosses, respectively. We follow the definition of Mach number as the ratio of CME speed to the sum of Alfv\'{e}n speed and solar wind speed. The Alfv\'{e}n speed estimation was produced using the magnetic field and plasma density models (\citealt{dulk}; \citealt{leblanc}; \citealt{mann}; \citealt{gopalswamy2001c}; \citealt{eselevich}).\\

 The Alfv\'{e}n speed is defined as
 \begin{equation}
    V_A = 2 \times 10^6 \rho^{-1/2} B \,\, (kms^{-1})
 ,\end{equation}where B is the magnetic field strength in gauss and rho is the number density in cm$^{-3}$. They can be determined by
 \begin{equation}
 B = 2.2 r^{-2} \,\, (G)
  \end{equation}
 and
 \begin{equation}
 \rho(r) = 3.3 \times 10^5 r^{-2} + 4.1 \times 10^6 r^{-4} + 8.0 \times 10^7 r^{-6} \,\, (cm^{-3})
  \end{equation}The solar wind profile was obtained from the empirical relation developed by \citet{sheeley},
 \begin{equation}
 V_{SW}^2 = v_a^2   [1 - e^{ - (r - r_1/) / r_a} ]
,\end{equation}where r$_1$ = 4.5 R$\odot$, r$_a$ = 15.2 R$\odot$, v$_a$ = 418.7 km\,s$^{-1}$, and r is the heliocentric distance in R$_{sun}$. The same procedure was repeated for the >50 MeV and >100 MeV energy channels.\\
 
  The protons in the considered energy range need propagation times of about 69 (10 MeV), 31 (50 MeV), and 22 (100 MeV) minutes to reach the Earth. This means that the slowest protons arrive one hour later than the light. Therefore, their detection is formally delayed by about an hour compared to the observations carried out by coronagraphs. The delay is about 10 minutes less because the peaks of SEPs are reached when the CMEs are at some distance from the Sun. However, for the consideration of the relationship between SEP flux peak and the maximum CME speed this problem is completely negligible. This effect can only be relevant to the correct determination of the speed of CME at SEP peak. However, as  can be seen in Figure~2 (red cross), this speed is determined at some distance from the Sun, where its change is very slow. The CME after reaching the maximum  velocity propagates at almost constant speed. In one hour the CME velocity can change by not more than 5$\%$. On the other hand, the error in determining the speed using a linear fit is about 15$\%$ (\citealt{Michalek17}). Hence, we have neglected this effect in our study.

   It is important to emphasize why we have tried to use the data of SOHO and STEREO to determine the speed. The 90$^{\circ}$ separation of the STEREO twin spacecraft with respect to the Earth is known as quadrature configuration. This position proves to be  advantageous as the halo CME kinematics can be observed more accurately. The speeds obtained by STEREO are closer to the spatial (real) speeds. The SOHO measurements on the other hand can be significantly modified by projection effects. We would like to compare the outcome of the results obtained by these two data to determine the correlations. Figure~3 shows the STEREO spacecraft configuration on the days of observation of the first (14 August 2013) and last (28 December 2010) CME studied in this paper. During these days the spacecraft were separated from the Earth by 80$^{\circ}$ and 150$^{\circ}$, respectively. Since very few energetic CMEs were recorded in this period, to obtain a sufficiently large number of events we had to extend our observations to the end of 2013. In 2013 the location of STEREO satellites differed significantly from quadrature. From our sample, a majority of the events originate at western longitudes and they happen after STEREO-A quadrature, therefore it can be expected that STEREO-A is the spacecraft that offers a better side view of the CME. On the other hand, it is worth noting that STEREO observations have a significant advantage in comparison with SOHO observations. In terms of field of view, STEREO/SECCHI offers a wider range of observation, 1.5 R$_{sun}$ - 318 R$_{sun}$, whereas the SOHO/LASCO C2/C3 field of view is 1.5 R$_{sun}$ - 30 R$_{sun}$. Hence, STEREO offers an advantage in studying the CME kinematics at large distances  from Sun, during peak SEP intensities in the heliosphere.


\begin{figure*}
   \centering
   \includegraphics[width=15cm,height=7cm]{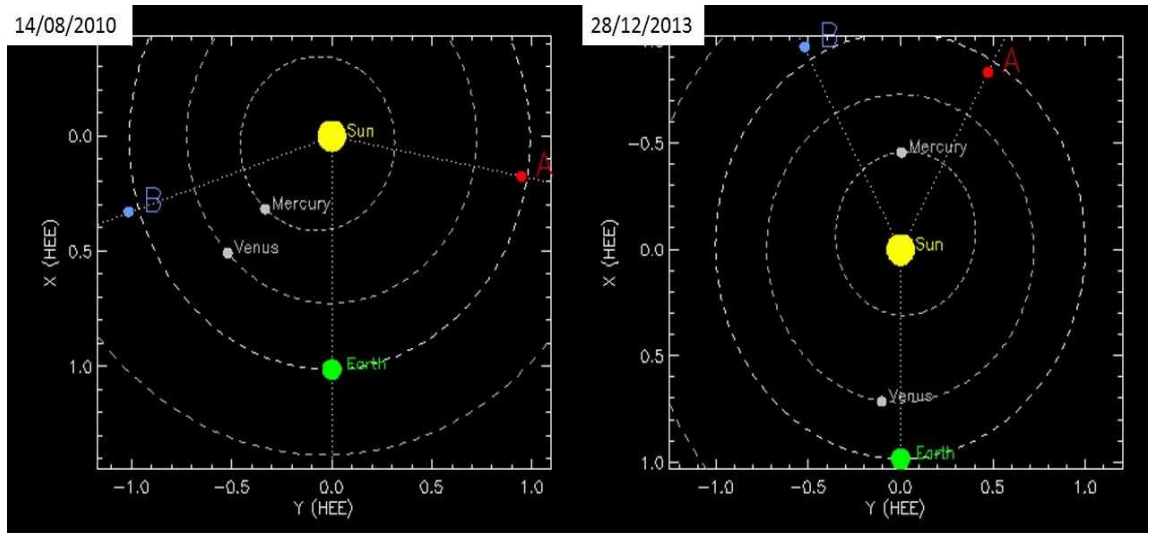}
   \caption{STEREO-A and -B configuration with respect to Earth on the days of observation of the first (14 August 2010) and
last (28 December 2013) CME studied in this paper. During these days the spacecraft had an angular separation with respect to the Earth of 80$^\circ$ and
150$^\circ$, respectively, marking the quadrature configuration. Images from STEREO orbit tool of the STEREO science center (https://stereo-ssc.nascom.nasa.gov/where/).}
              \label{FigGam}%
    \end{figure*}

   \begin{figure*}
   \centering
   \includegraphics[width=18cm,height=10cm]{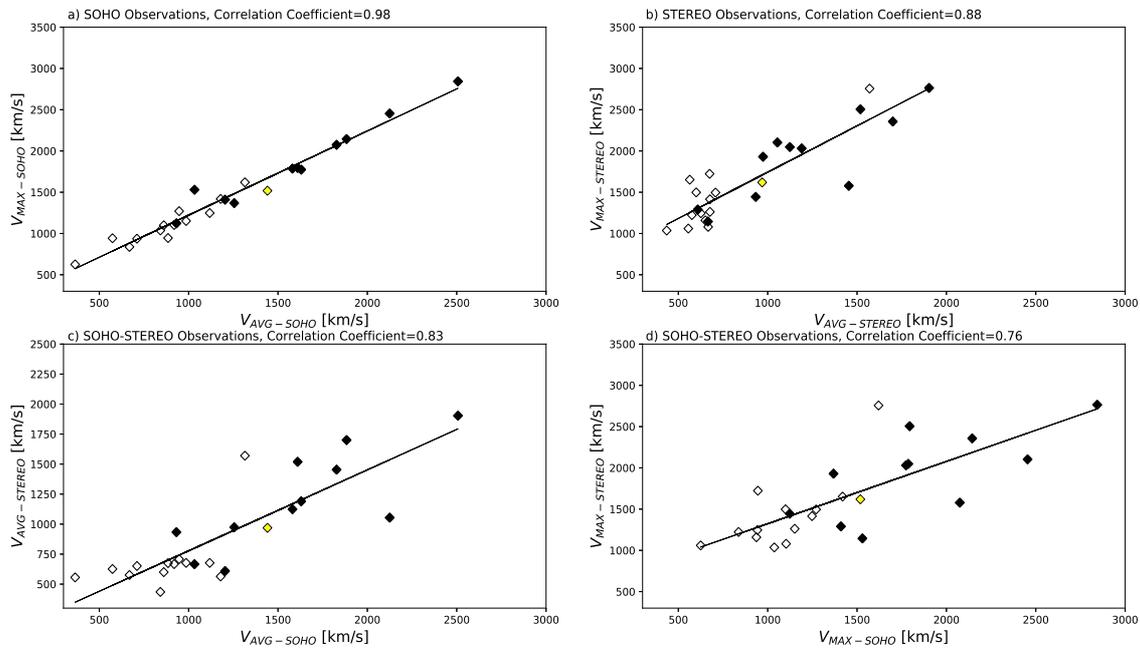}
   \caption{Relationship between (a) the average and maximum velocity determined in  the SOHO images, (b) the average and maximum velocity determined in the STEREO images, (c) the average velocity determined in SOHO and STEREO images, (d) the maximum velocity determined in SOHO and STEREO images. The open symbols represent disk events (longitude -20<L<45) and filled symbols represent west events (Longitude >45). There is one east event, on 31 August 2012 (longitude <-20;  in yellow). This is the only event in our sample analyzed by STEREO-B as it exhibited good quality.}
              \label{FigGam}%
    \end{figure*}
    
    \begin{figure*}[!h]
   \centering
   \includegraphics[width=16cm,height=10cm]{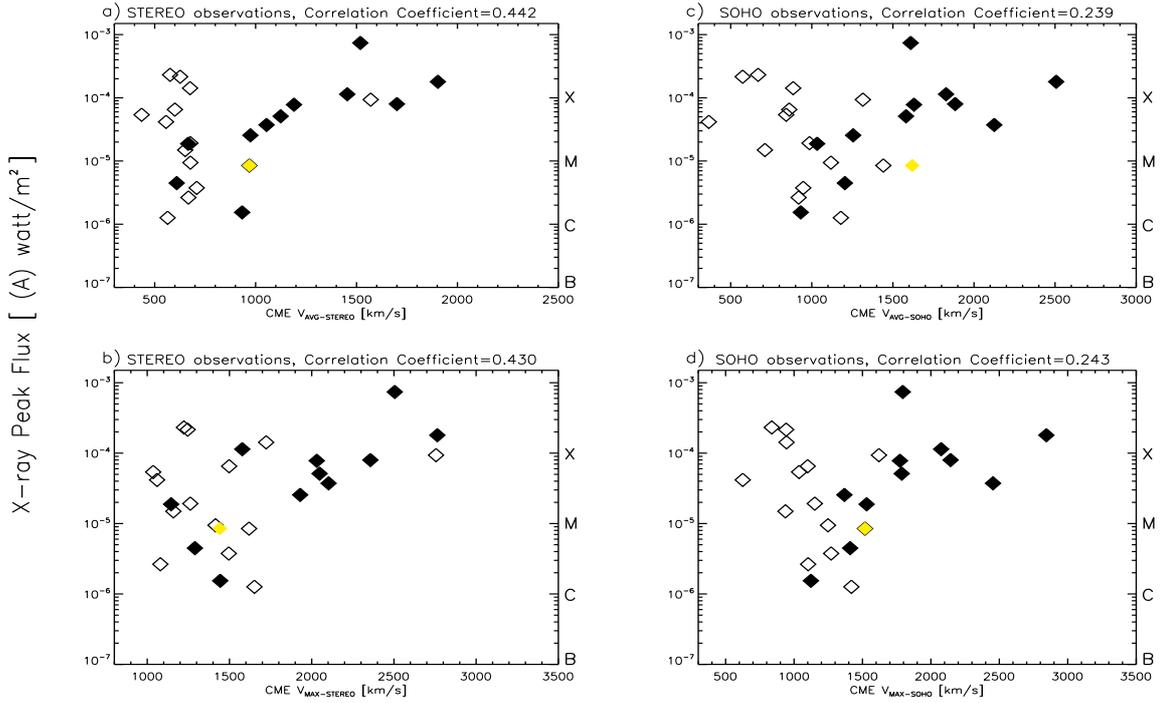}
   \caption{Scatter plots of X-ray peak fluxes of solar flares (0.1-0.8 nm) vs respective initial velocity of CMEs: (a)   average velocity-STEREO, (b)   maximum velocity-STEREO, (c)   average velocity-SOHO, (d)   maximum velocity-SOHO. The open symbols represent disk events (longitude -20<L<45) and filled symbols represent west events (Longitude >45). There is one east event,  on 31 August 2012 (longitude <-20; in yellow). This is the only event in our sample analyzed by STEREO-B as it exhibited good quality. }
              \label{FigGam}%
    \end{figure*}
    
   \begin{figure*}[!h]
   \centering
   \includegraphics[width=16cm,height=10cm]{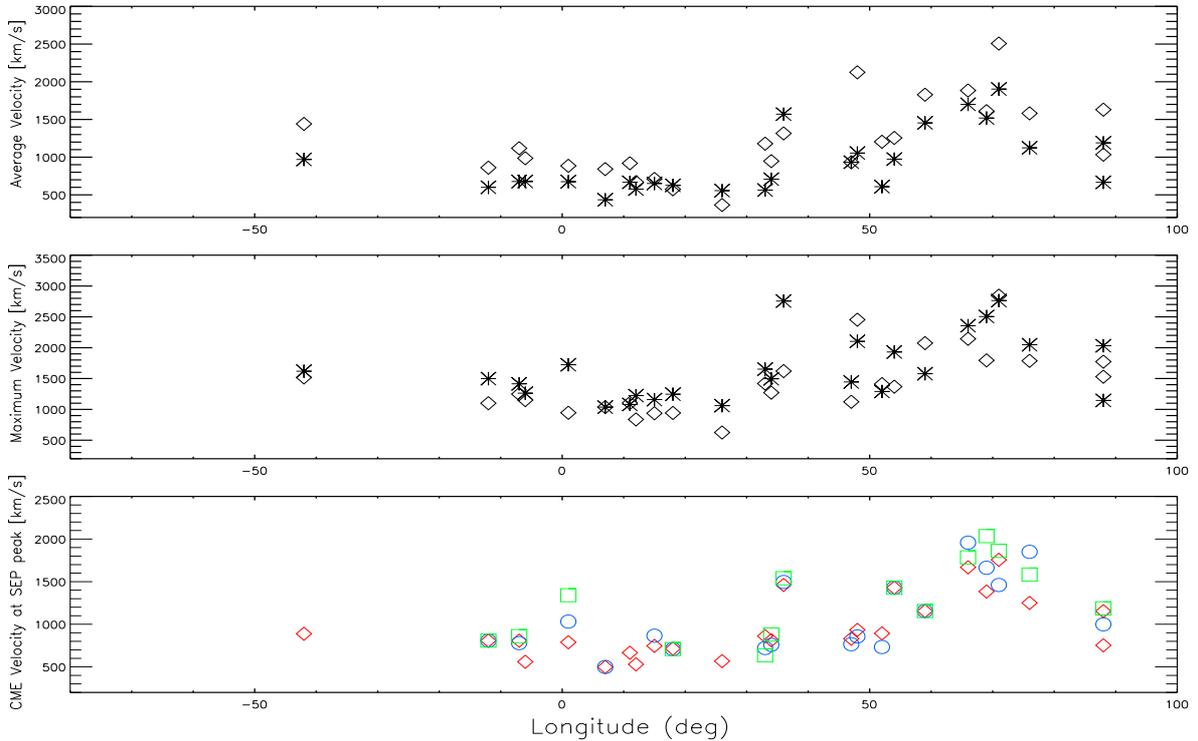}
   \caption{Scatter plots of average velocity, maximum velocity, and CME velocity at SEP peak flux vs  longitude of the active regions associated with the respective CMEs. In the two upper panels, diamonds and stars show velocities for SOHO and STEREO, respectively. In the bottom panel,
  colors are assigned to the SEP peak flux in the energy channels: >10 MeV (red), >50 MeV (blue), and >100 MeV (green).}
              \label{FigGam}%
    \end{figure*}
    
    \begin{figure*}
   \centering
   \includegraphics[width=16cm,height=10cm]{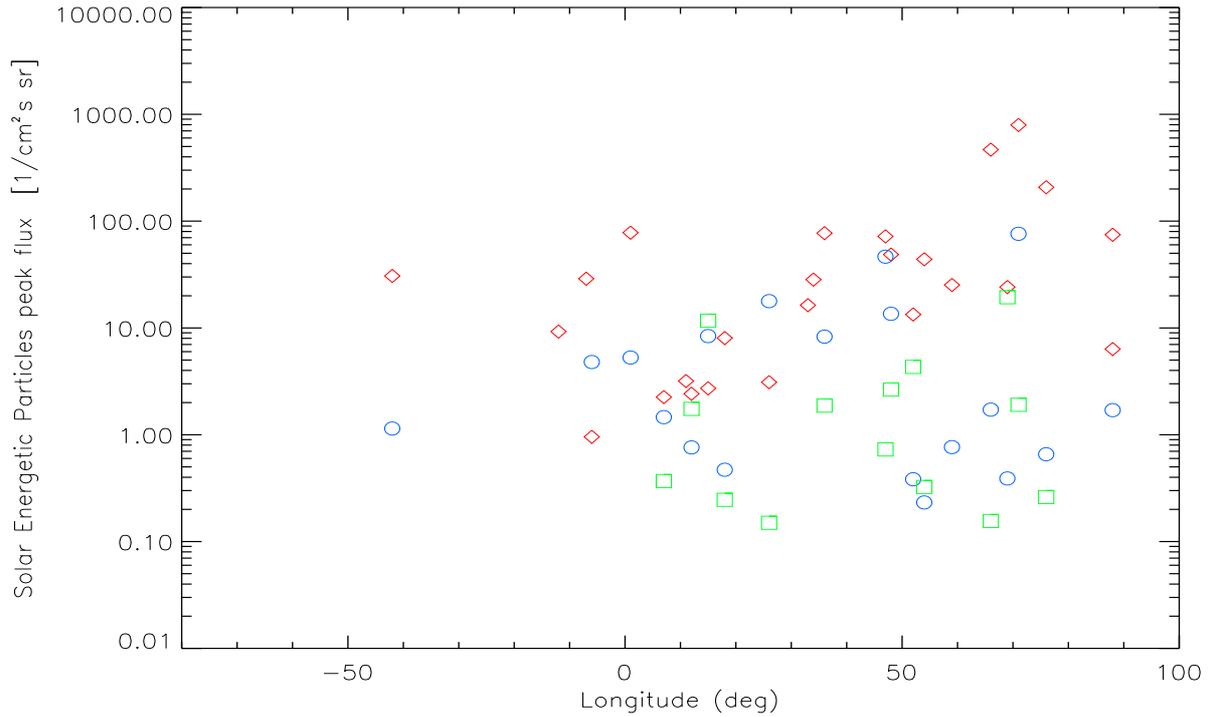}
   \caption{Scatter plots of SEP peak flux vs longitude of the active regions associated with the respective CMEs. Colors are assigned to the SEP peak flux in the energy channels: >10 MeV (red), >50 MeV (blue), and >100 MeV (green).}
              \label{FigGam}%
    \end{figure*}

   \begin{sidewaystable*}[!h]
\caption{Observational parameters of 25 CMEs and the associated SEPs and Flares during the period 2009-2013.}\label{YSOtable}
\centering
\tiny
\begingroup
\setlength{\tabcolsep}{5pt} 
\renewcommand{\arraystretch}{1.8} 

\begin{tabular}{ |l| c c c| c c| c c c| c c c| c| c c c| c c| }
\hline
\hline
\multicolumn{4}{|c|}{SOHO CME [km/s] }&\multicolumn{2}{c|}{STEREO CME [km/s] }&\multicolumn{3}{c|}{Peak proton flux (pfu)}&\multicolumn{3}{c|}{CME speed at peak proton flux [km/s]}&\multicolumn{1}{c|}{Mach at }&\multicolumn{3}{c|}{Mach at peak proton flux}&\multicolumn{2}{c|}{Solar Flare}\\

\hline
\# & Date \& Time & V$_{AVG}$ & V$_{MAX}$ & V$_{AVG}$ & V$_{MAX}$ & >10 MeV & >50 MeV & >100 MeV & >10 MeV & >50 MeV & >100 MeV & V$_{MAX}$ & >10 MeV & >50 MeV & >100 MeV  & Class & Location\\
\hline

  1 & 20100814 10:12 & 1204 & 1409 & 609  & 1290 & 13.371 & 0.3829 & -      & 893  & 733  & -    & 1.04  & 1.08 & 0.76 & -    & C4.4 & N17W52 \\
  2 & 20110215 02:24 & 669  & 837  & 576  & 1223 & 2.4200 & -      & -      & 530  & -    & -    & 0.97  & 0.88 & -    & -    & X2.2 & S20W12 \\
  3 & 20110307 20:00 & 2125 & 2454 & 1054 & 2103 & 48.665 & 0.7609 & -      & 934  & 857  & -    & 1.73  & 1.83 & 1.67 & -    & M3.7 & N30W48 \\
  4 & 20110607 06:49 & 1255 & 1368 & 974  & 1930 & 43.959 & 13.556 & 4.3137 & 1429 & 1429 & 1429 & 1.37  & 2.27 & 2.27 & 2.27 & M2.5 & S21W54 \\
  5 & 20110802 06:36 & 711  & 936  & 651  & 1159 & 2.7207 & 0.2323 & -      & 747  & 866  & -    & 1.53  & 1.29 & 1.24 & -    & M1.4 & N14W15 \\
  6 & 20110804 04:12 & 1315 & 1620 & 1570 & 2756 & 77.155 & 8.3902 & 1.7458 & 1461 & 1496 & 1540 & 2.35  & 2.76 & 2.63 & 2.63 & M9.3 & N19W36 \\
  7 & 20110809 08:12 & 1610 & 1794 & 1519 & 2505 & 24.123 & 8.2908 & 2.6508 & 1384 & 1663 & 2035 & 2.03  & 2.31 & 1.76 & 1.92 & X6.9 & N17W69 \\
  8 & 20110906 02:24 & 842  & 1037 & 435  & 1036 & 2.2524 & 0.3895 & -      & 499  & 499  & -    & 0.66  & 0.67 & 0.67 & -    & M5.3 & N14W07 \\
  9 & 20110906 23:05 & 574  & 942  & 626  & 1246 & 8.0393 & 1.4613 & 0.3242 & 714  & 709  & 709  & 1.60  & 1.22 & 1.11 & 1.11 & X2.1 & N14W18 \\
  10 & 20111126 07:12 & 932 & 1123 & 933  & 1444 & 72.034 & 0.4702 & -      & 831  & 766  & -    & 1.86  & 1.63 & 1.44 & -    & M4.0 & N11W47 \\
  11 & 20111225 18:48 & 365 & 625  & 556  & 1060 & 3.1043 & -      & -      & 568  & -    & -    & 1.44  & 0.89 & -    & -    & M4.0 & S22W26 \\
  12 & 20120127 18:27 & 2507 & 2843 & 1903 & 2764 & 794.04 & 46.492 & 11.699 & 1756 & 1461 & 1861 & 2.85 & 3.36 & 2.69 & 3.14 & X1.7 & N27W71 \\
  13 & 20120313 17:36 & 1884 & 2144 & 1700 & 2357 & 467.56 & 17.826 & 1.8739 & 1668 & 1958 & 1785 & 1.74 & 2.72 & 2.71 & 2.63 & M7.9 & N17W66 \\
  14 & 20120517 01:48 & 1581 & 1786 & 1124 & 2048 & 207.56 & 76.123 & 19.386 & 1251 & 1850 & 1583 & 1.24 & 1.95 & 2.27 & 2.09 & M5.1 & N11W76 \\
  15 & 20120614 14:12 & 986  & 1151 & 677  & 1262 & 0.9552 & -      & -      & 560  & -    & -    & 1.35 & 1.02 & -    & -    & M1.9 & S17E06 \\
  16 & 20120706 23:24 & 1828 & 2074 & 1454 & 1578 & 25.241 & 1.7187 & 0.3686 & 1155 & 1155 & 1155 & 2.36 & 1.92 & 1.92 & 1.92 & X1.1 & S13W59 \\
  17 & 20120712 16:48 & 885  & 945  & 675  & 1723 & 77.968 & 0.6569 & 0.2459 & 789  & 1032 & 1341 & 1.25 & 1.29 & 1.43 & 1.38 & X1.4 & S15W01 \\
  18 & 20120719 05:24 & 1630 & 1773 & 1190 & 2031 & 74.575 & 4.7952 & 0.7291 & 1154 & 999  & 1186 & 2.35 & 2.16 & 1.77 & 1.93 & M7.7 & S13W88 \\
  19 & 20120831 20:00 & 1441 & 1518 & 969 & 1619 & 30.671 & -       & -     & 889   & -    & -    & 1.21 & 1.85 & -    & -    & C8.4 & S19E42 \\
  20 & 20120928 00:12 & 947  & 1270 & 708 & 1496 & 28.363 & 0.7662 & 0.1498 & 814   & 759  & 880  & 1.26 & 1.31 & 1.25 & 1.38 & C3.7 & S06W34 \\
  21 & 20130411 07:24 & 861  & 1099 & 600 & 1498 & 9.2449 & 5.2834 & 1.9079 & 808   & 808  & 808  & 1.04 & 1.12 & 1.01 & 1.01 & M6.5 & N07E12 \\
  22 & 20130421 07:24 & 919  & 1103 & 667 & 1080 & 3.1845 & -      & -      & 667   & -    & -    & 0.66 & 1.06 & -    & -    & B8.7 & N14W11 \\
  23 & 20130929 22:12 & 1179 & 1419 & 564 & 1652 & 16.315 & 1.6996 & 0.1556 & 861   & 721  & 635  & 1.24 & 1.49 & 1.28 & 1.16 & C1.2 & N10W33 \\
  24 & 20131107 00:00 & 1033 & 1530 & 666 & 1145 & 6.3445 & -      & -      & 753   & -    & -    & 1.09 & 1.17 & -    & -    & M1.8 & S11W88 \\
  25 & 20131228 17:36 & 1118 & 1247 & 677 & 1415 & 28.900 & 1.1433 & 0.2608 & 808   & 779   & 861 & 1.14 & 1.30 & 1.20 & 1.24 & C9.3 & S18E07 \\
\hline\hline

\end{tabular}
\endgroup
\end{sidewaystable*}

Our considerations are summarized in  Table~1. In the first four columns we have date, time, average and the maximum velocities of a given CME taken from the SOHO/LASCO catalog. In Cols. 5 and 6 we show the average and the maximum velocities obtained from STEREO images. Columns 7-9 shows the peak SEP fluxes in the three energy channels. The next three columns present CME speeds at peak SEP fluxes for these energy channels. Column 13 gives the Mach number at maximum CME velocity for STEREO observations. Columns 14-16 show Mach numbers at peak SEP fluxes for the considered energy channels. We determined Mach number only for instantaneous velocities of CME by STEREO. The last two columns present the class and location of solar flares associated with CMEs from GOES data.

\section{Analysis and results}
   Our study mainly concentrates on the influence of CME speed on the production of SEPs. Specifically, we wanted to know the statistics of non-interacting CMEs and the associated SEPs. Though there were several weak, minor and major SEP events during the ascending phase of solar cycle 24, we took into consideration only those that had significant intensity above the background flux value, i.e., $\geq$1 pfu for the >10 MeV energy band, $\geq$0.1 pfu for the >50 MeV and >100 MeV energy bands. We utilized both SOHO/LASCO and STEREO/SECCHI data to study the kinematics of the CMEs but the quadrature configuration of STEREO during 2009 - 2013 could provide true radial speeds. Results of the study are presented in following sections.

\subsection{CME kinematics}

  The non-interacting CMEs had an average speed range of 365/435 km~s{$^{-1}$} to 2507/1903 km~s{$^{-1}$} and maximum speed range of 625/1036 km~s{$^{-1}$} to 2843/2764 km~s{$^{-1}$} for SOHO and STEREO, respectively. Since they are all halo CMEs their sky-plane widths, as defined by the LASCO CME catalog, are 360$^{\circ}$.

Relationships between the considered velocities are presented in Figure~4.
  In Figure 4, panels a and b   show the relationships between the average and maximum speeds determined in the SOHO images. A similar relationship for the STEREO images is shown in panels c and d. Solid lines are linear fits to data points. It can be seen that the average ejection velocities are strongly correlated with their maximum velocities (panels a and b) regardless of the instrument used to determine them. An almost perfect correlation, which is the result of a limited field of view of the SOHO instruments, appears for the SOHO observations (panel a).
  We can see that the maximum velocities are much higher, on average 79.4\%, than the average velocities in the case of observations from the STEREO spacecraft (panel b). For the SOHO observations, the maximum velocities are on average only
 18.6\% higher than the average velocities. This  result is due to the field of view of the
STEREO instruments used to determine the velocity profiles (COR1, COR2, HI1, and HI2), which is much
larger than the field of view of the LASCO coronagraphs (C2 + C3). It means that the field
of view of the STEREO telescopes covers the area where CMEs undergo significant deceleration due to interaction with the solar wind. For this reason, the average velocities of the CMEs determined from the STEREO observations are significantly lower than the other
speeds determined in these studies. 
We think it important to note that the determination of instantaneous CME velocities at large distances from the Sun has no direct relationship with the search for maximum velocities or velocities at SEP peak, which are determined relatively close to the Sun. However, fluxes of energetic particles are produced during the entire CME passage to the Earth, so it is also important to determine their velocities during the same distance, if possible. In addition, including the SECCHI Heliospheric Imager field of view to large heliocentric distances allowed us to show that instantaneous CME velocities change radically in time and space. In addition,  the comparison of average velocities obtained from SOHO and STEREO clearly show that the values of these velocities significantly depend on the instrument with which they were determined, i.e., number of height--time points. The velocities obtained in this way must be treated with some caution.

Correlations between the speeds for these two instruments are slightly smaller. The correlation coefficients are respectively 0.83 and 0.76  for the average and maximum speeds.
In this case the wider dispersion of speeds is the result of being  determined
from  two different spacecraft (SOHO and STEREO) that observe the Sun at different positions and fields of view. For the same reason as described in the paragraph above, the average speeds for SOHO are  31.7\% higher and the maximum speeds are  8.7\% lower than those registered for STEREO.

\begin{figure*}
   \centering
   \includegraphics[width=18cm,height=10cm]{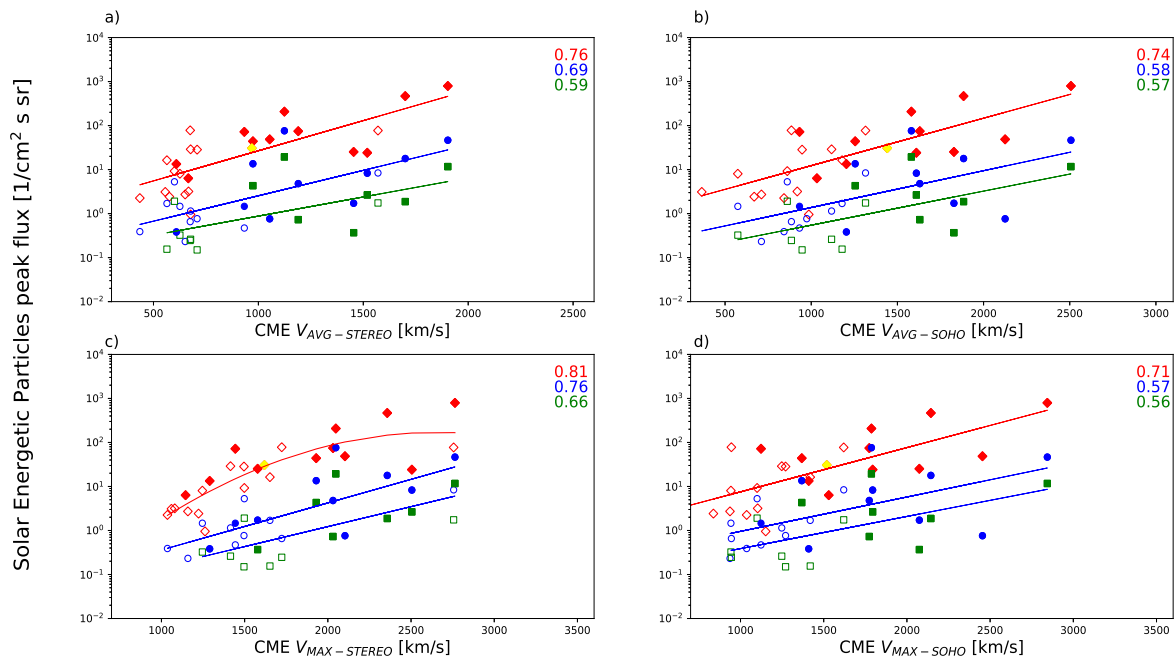}
   \caption{Scatter plots of maximum of SEP flux at three energy bands (>10 MeV (red), >50 MeV (blue), and >100 MeV (green)) vs the respective  speed of CMEs (average speed-STEREO (panel a), maximum speed-STEREO (panel b), average speed-SOHO (panel c), maximum speed-SOHO (panel d) ). The open symbols represent disk events (longitude -20<L<45) and filled symbols represent west events (Longitude >45).  There is one east event, on 31 August 2012 (longitude <-20; in yellow). This is the only event in our sample analyzed by STEREO-B as it exhibited good quality. The solid curves are linear fits to the data points.}
              \label{FigGam}%
    \end{figure*}

 Figure~5 shows relations between respective speeds of CME and X-ray peak fluxes of the associated solar flares in the wavelength range 0.1-0.8 nm.
The left panels present results for STEREO and the right panels for SOHO. Additionally, we separated the sample into disk and limb events according to their longitude. We see that the intensities of X-ray flares are not correlated with speeds recorded in the SOHO/LASCO coronagraphs. Correlation coefficients are poor (<0.25). Due to projection effects, the disk events are mostly shifted to the left. This shows that, on average, disk events have much lower velocities in comparison to the limb events.  For STEREO observations the results are different (panels a and b). Correlation coefficients are significantly larger (>0.4) and velocities of disk events are less scattered in STEREO than SOHO at wider velocity ranges.

 \subsection{Location and SEP events}
Figure~6 displays  scatter plots of average velocity (panel a), maximum velocity (panel b), and velocity at SEP peak flux (panel c) versus longitude of location of X-ray flare associated with respective CMEs. In the two upper panels, diamonds and stars show velocities for SOHO and STEREO, respectively. In the bottom panel,   colors are assigned to SEP peak flux in the energy channels (>10 MeV (red), >50 MeV (blue) and >100 MeV (green). This selection of colors is kept throughout the paper. It is interesting that for the all the considered velocities (for the three panels) we can distinguish three different ranges of longitudes. CMEs originating close to the disk center (longitude<35$^\circ$) have comparatively low velocities. For events originating in the longitude range between 35$^\circ$ and 65$^\circ$ we observe a significant increase for the all the considered velocities. Since the increase in velocity does not depend on instrument and is rather an estimated velocity, we can assume that the projection effect is less significant for ejections moving at angle
$\approx45^\circ$ relative to the observing instrument. For limb events (longitude> 90$^\circ$ ) the velocities decrease to values observed for the disk center events. This conclusion is proved by the results presented in Figure~7. In this figure we show a scatter plot of SEP peak flux versus longitude of X-ray flare. We do not see any clear variation in  intensities of SEP with location of active regions on the Sun. The disk and west limb events can produce SEP events with fluxes $\approx$100 pfu in the 10 MeV energy channel.

\begin{figure*}
   \centering
   \includegraphics[width=18cm,height=10cm]{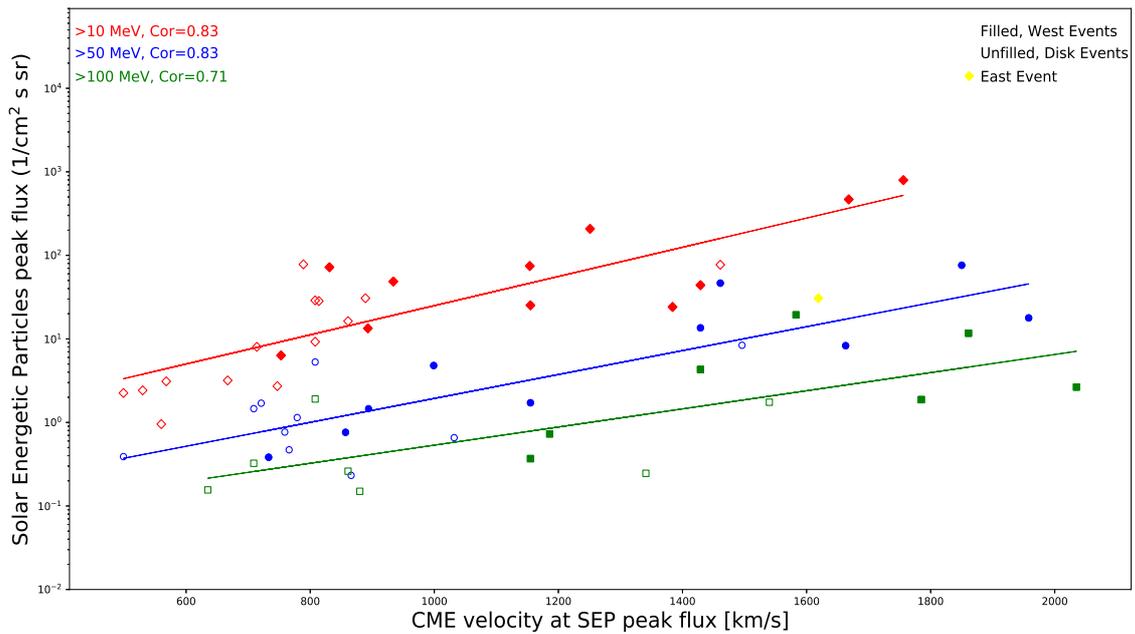}
   \caption{Scatter plot of CME speed observed at the peak SEP flux vs SEP flux at three energy bands: >10 MeV (red), >50 MeV (blue) and >100 MeV (green). The open symbols represent disk events (longitude -20<L<45) and filled symbols represent west events (Longitude >45).  There is  one east event,  on 31 August 2012 (longitude <-20; in yellow). This is the only event in our sample analyzed by STEREO-B as it exhibited good quality.}
              \label{FigGam}%
    \end{figure*}

\subsection{Velocity and SEP intensities}
In the present study, our main aim is to determine the relationship between CME speeds and SEP intensities. The motivation is derived from the previous studies where it is shown that CME speeds mostly affect SEP intensities. Figure~8 shows
scatter plots of SEP peak fluxes in the three energy channels versus the average and maximum speeds of CMEs. The left panels (a and c) present results for STEREO and the right panels (b and d) for SOHO observations. The linear fits are suitable for almost all the scatter plots except for the CME maximum velocity versus the >10 MeV SEP peak flux in panel c. The quadratic fit represents the profile variation correctly. The diagrams show a few interesting results. The considered velocities are well correlated with SEP peak fluxes. The CME velocities are the best correlated (from 0.716 up to 0.810) with SEP fluxes in the >10 MeV energy channel (in red). The poorest correlations are for the most energetic protons in the >100 MeV energy channel (in green) from 0.562 up to 0.665). Correlations between the velocities of CME and maximum fluxes of protons are poorer for SOHO observations than for STEREO data. They are similar for the average and maximum speeds.
In the case of STEREO data we observe significant correlation between the maximum speed and the  intensitiy of protons. Correlation coefficients are equal to 0.810, 0.756, and 0.665 for protons at energy channels >10, >50, and >100 MeV, respectively. The respective correlation coefficients are about 0.05 larger than those obtained from STEREO data for the average velocities.

During the gradual SEP events energetic particles are produced continuously by shocks generated by fast CMEs moving in the interplanetary space. At the time when the velocity of the CMEs reaches the Mach number in the solar wind, they start to produce energetic particles. From that moment we observe a systematic increase in fluxes of energetic particles. At some point, the intensities of the particles reach their maximum values, which depends on two factors: speed of ejection of CME and their magnetic connectivity with the Earth. The best nominal magnetic connectivity with the Earth corresponds to a CME originating around W60 on the solar disk. For these bursts, the energetic particles reach the  maximum flux quickly after their eruption from the Sun. In the case of ejections located in the center of the solar disk or on its eastern part, which are not well connected magnetically to the Earth, the maximum intensity of energetic particles is achieved much later, i.e., when ejections expand enough so that their fronts are well connected magnetically to the Earth. At the same time, when the ejection expands, its speed decreases. This means that when we observe the maximum intensities of energetic particles, the ejection speed may be much lower than their maximum value. This can be clearly observed in Figure 2. It is worth noting that while analyzing the time of maximum SEP fluxes, the interplanetary scattering effects are not taken into consideration. In addition, the possible influence of the angular separation between the magnetic footpoint of the observing spacecraft and the source active region is not inspected. In this context, the instantaneous ejection speeds recorded at the time of maximum particle energy fluxes should be the better indicator of fluxes of SEP events than their maximum velocities. Therefore,  scatter plot of CME speeds observed at the SEP peak fluxes versus SEP peak fluxes are presented in Figure~9. For all three energy channels we noted significant correlation between these parameters (0.835, 0.831, and 0.719 for the >10, >50, and >100 MeV energy channels, respectively). This clearly indicates that instantaneous velocity at the peak of proton flux is the best indicator of intensities of particle events. Additionally, for STEREO observation we determined Mach numbers for the respective instantaneous velocities. Mach number is one of the most significant parameters determining the efficiency of acceleration of particles in the shock vicinity  (\citealt{Li2012a}, \citealt{Li2012b}).\\\\ 
	Figure~10 shows scatter plots of CME Mach number at the SEP peak flux (left panel) and CME Mach number at the maximum CME speed (right panel) versus maximum of SEP flux for the three energy channels. From the figure we see that SEP fluxes are significantly correlated with CME Mach number obtained when particle fluxes reach maximum values. The correlation coefficients are similar  to those obtained for velocities at the  SEP peak fluxes (0.835, 0.831, and 0.719 for >10, >50, and >100 MeV energy channels, respectively). The CME Mach number obtained at the maximum CME speeds are less correlated with proton intensities. It is important to notice that CME Mach numbers are obtained from the theoretical model and only approximate the true values.

\begin{figure*}
   \centering
   \includegraphics[width=9.1cm,height=6cm]{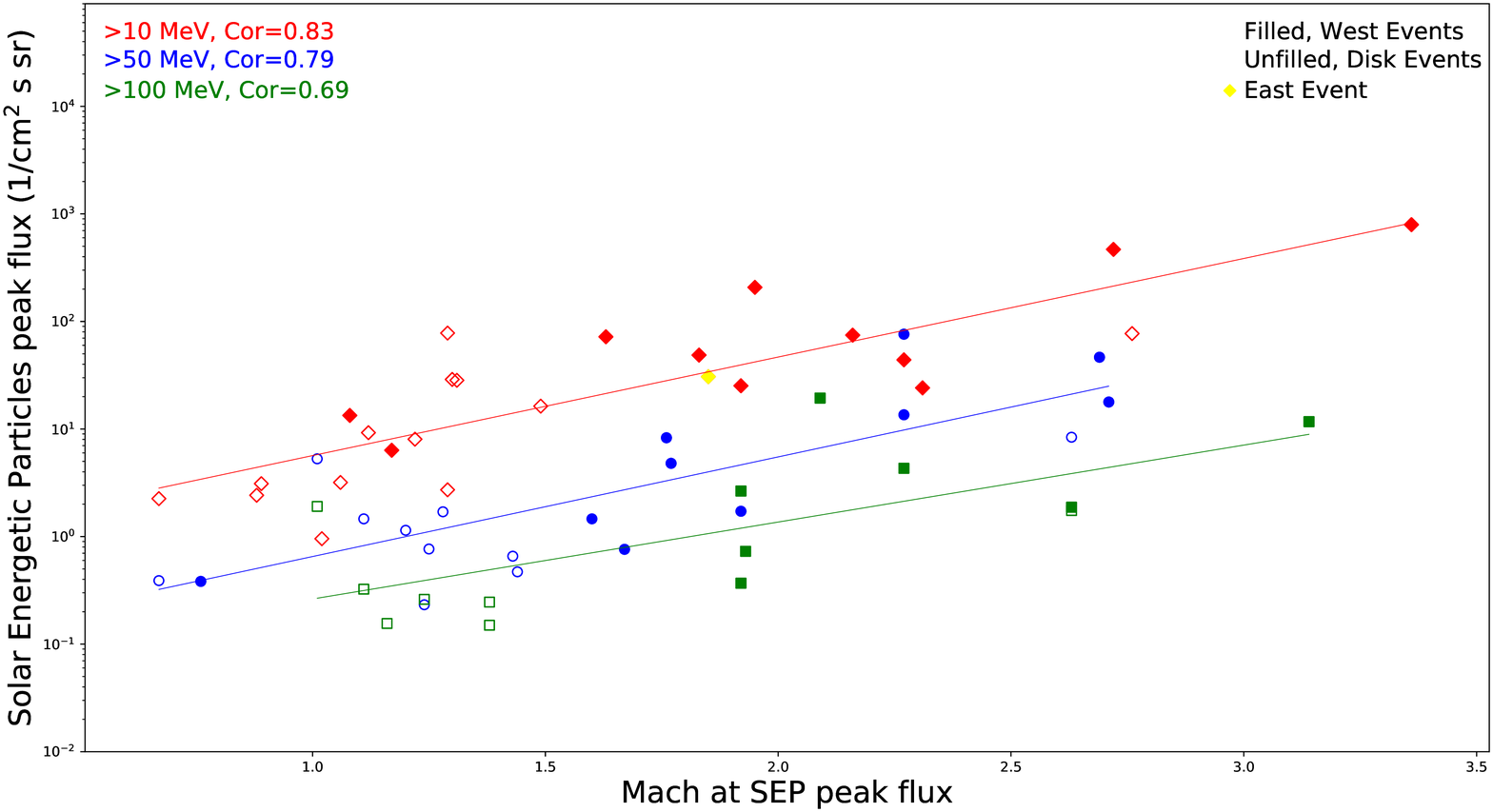}
   \includegraphics[width=9.1cm,height=6cm]{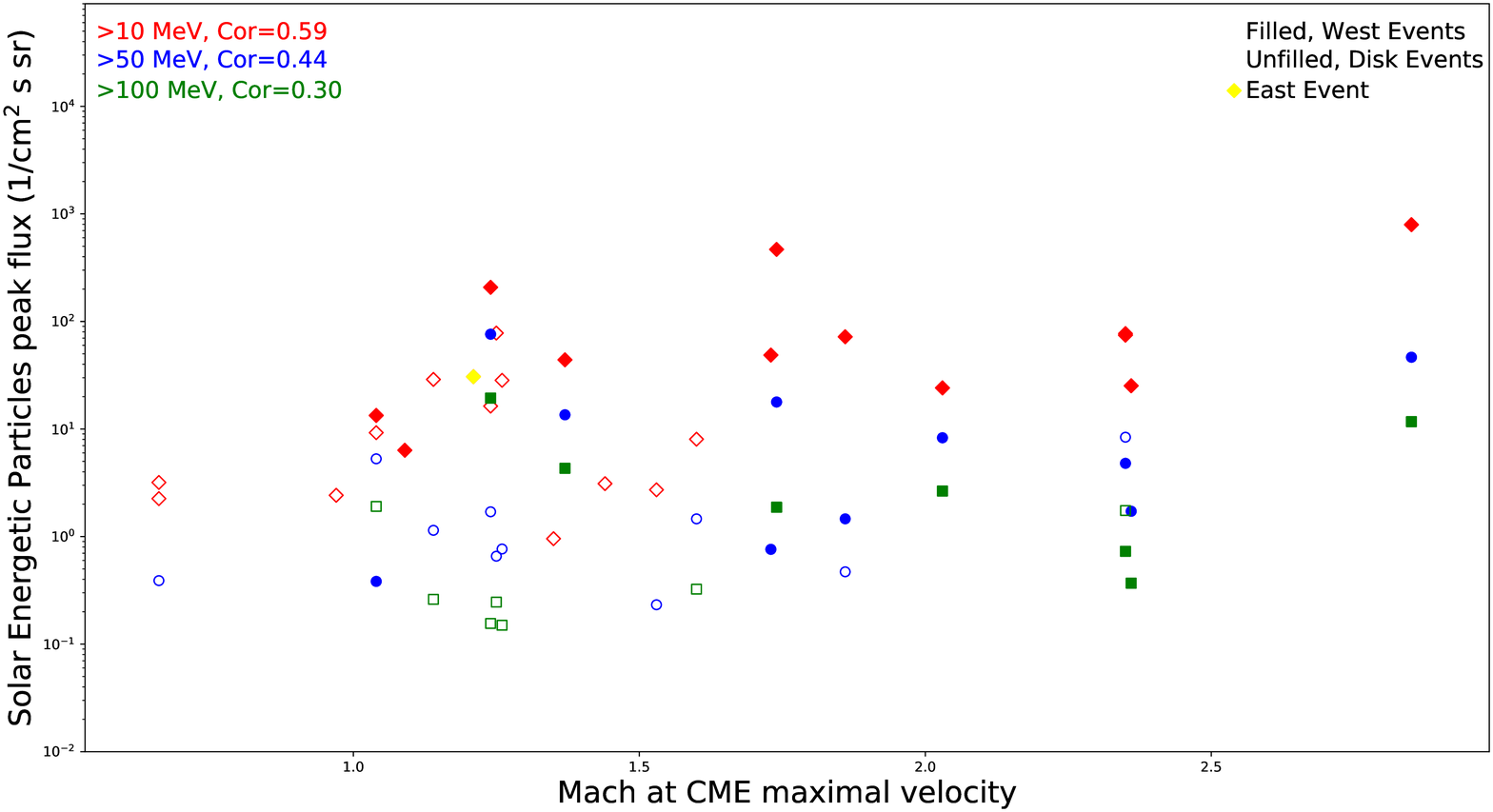}
   \caption{Scatter plot of CME Mach number observed at the peak SEP flux (left panel) and CME Mach number at the maximum CME speed (right panel) vs SEP peak flux in the >10 MeV (red), >50 MeV (blue), and >100 MeV (green) energy channels. The open symbols represent disk events (longitude -20<L<45) and filled symbols represent west events (Longitude >45).  There is  one east event, on 31 August 2012 (longitude <-20; in yellow).  This is the only event in our sample analyzed by STEREO-B as it exhibited good quality.}
              \label{FigGam}%
    \end{figure*}
    
\subsection{Associated flares and SEP events}
 As mentioned earlier, energetic particles can be generated by flares or shock waves generated by CMEs. It is therefore necessary to examine whether flares associated with CMEs affect the intensity of the observed protons. Figure~11 shows a scatter plot of X-ray peak flux versus SEP peak flux. We do not find any significant correlation between these parameters. This means that the flares do not have a significant effect on SEP intensities. Correlation coefficients are <0.3 for the three energy channels. However, it is important to note that intensities in the >10 MeV energy channel are predominantly accelerated by the CME shock (\citealt{Xie19}). They are most significantly correlated with instantaneous CME velocities, but completely uncorrelated with intensities of X-ray flares. The situation is a little different in the case of the higher energy protons of the >50 and >100 MeV energy channels. These show lower correlation, as was shown in the previous paragraph, with instantaneous CME velocities compared to >10 MeV protons. From Figure 11, for high energy protons of >50 and >100 MeV, we  find a poor correlation with the intensity of X-ray flares. In addition, we sometimes observe that the peaks of SEP particles in the >50 and >100 MeV energy channels appear earlier than the CME peak velocity and >10 MeV SEP peak flux. These observations may indicate that the >50 and >100 MeV protons are mostly accelerated by shock waves generated by fast CME, but flares may be partly involved in their acceleration process as well, especially in the first phase  of their ejections producing seed particles.

  \begin{figure*}[h!]
   \centering
   \includegraphics[width=16cm,height=10cm]{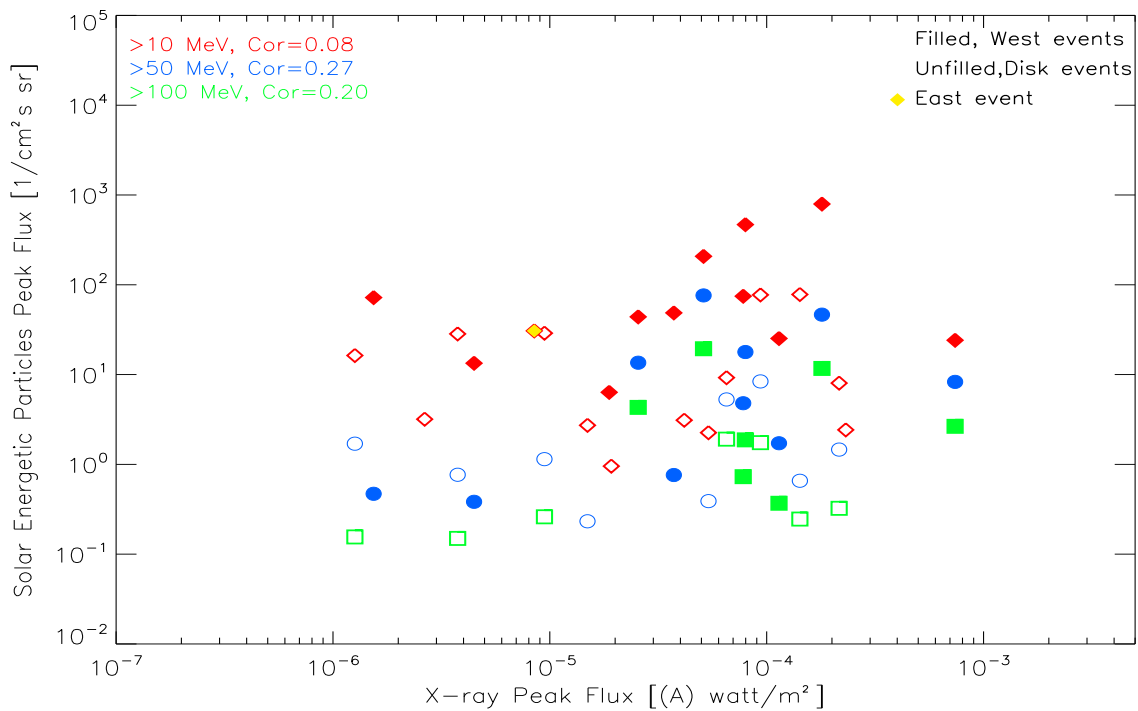}
   \caption{Scatter plot of X-ray peak flux vs SEP peak flux in the >10 MeV (red), >50 MeV (blue), and >100 MeV (green) energy channels. The open symbols represent disk events (longitude -20<L<45) and filled symbols represent west events (Longitude >45).  There is  one east event, on 31 August 2012 (longitude <-20;  in yellow). This is the only event in our sample analyzed by STEREO-B as it exhibited good quality.}
              \label{FigGam}%
    \end{figure*}

\section{Conclusions}

 We studied the kinematics of 25 CMEs and their associated SEPs and X-rays during the ascending phase of Solar Cycle 24. The CME data were obtained from instruments on board two separate spacecraft, SOHO/LASCO and STEREO/SECCHI. We chose the ascending phase of Solar Cycle 24 (2009-2013) because the STEREO twin spacecraft were near quadrature configuration with respect to the Earth. This position offered a big advantage in the accurate determination of the plane-of-sky speed, which is close to the true radial speed of halo events. We were also able to compare results obtained from the two different points of view of SOHO/LASCO and STEREO/SECCHI.
 Additionally, STEREO/SECCHI offers a wider range of observation, 1.5~R$_{sun}$--318~$_{sun}$, whereas the SOHO/LASCO C2/C3 field of view is 1.5~$_{sun}$--30~R$_{sun}$. Hence, STEREO offers an advantage in studying the CME kinematics at large distances  from Sun, during peak SEP intensities in the heliosphere. In the study we employed a new approach. From the height--time points we determined variations of CME speed in time and distance from the Sun. With  these velocity profiles we were able to determine CME speeds at any time during their expansion in the interplanetary space. For SOHO observations, we used the height--time points from the SOHO/LASCO catalog, and for the STEREO/SECCHI observations we determined the height--time points manually. To determine the instantaneous velocity profiles we used linear fits to five height–time data points. Shifting successively these five-point linear fits, we obtained the instantaneous profiles  of speed in time and in distance from the Sun. They formed the basis of our research. Using our method we selected 25 non-interacting CMEs associated with SEP events >10, >50 and >100 MeV energy channels. Their properties are summarized in Table~1.

 Analyzing the observational data, we obtained the following interesting results.

\begin{itemize}
  \item As is shown in Figure~4, the considered velocities are well correlated. Due to different positions and field of view they  differ slightly between instruments.\\

 \item The CME maximum and average velocities obtained by STEREO/SECCHI are moderately correlated with intensities of associated X-ray flares (Fig.~5). We do not observe this correlation for the data obtained from SOHO.\\

 \item The results also show that the considered velocities increase for events having longitudes in the range from 35$^{\circ}$  to 65$^{\circ}$. This seems to be due to a decrease in projection effects.\\

 \item The focus of the research was to determine the correlation between instantaneous CME speeds and intensities of energetic particles. We demonstrated that all the considered CME velocities are well correlated with intensities of energetic proton for the three energy channels. However, the most significant correlation, as we expected, was recorded for velocities determined at the SEP peak fluxes. In this case, the correlation coefficient reached 0.835 for the >10 MeV energy channel. Similar results were obtained in the case of CME Mach numbers determined  at the SEP peak fluxes.\\

 \item The good correlation obtained for fluxes of protons in the >10 MeV energy channel supports the hypothesis that the protons are predominantly accelerated by shock waves generated by fast CMEs propagating in the interplanetary space. The >50 and >100 MeV energetic particles are also mostly accelerated by the same shock waves, but a partial contribution by the associated flares can be also involved in their acceleration.
\end{itemize}

Our results support the hypothesis that energetic particles observed in the considered energy range are predominantly accelerated by CME shock. Nevertheless, as stated by Miteva et al (2013),
the correlation between SEP intensities and CME speed or X-ray intensity is not sufficient to completely deny combined participation of both CME-related and flare-related acceleration due to the interdependence of the two phenomena.

\begin{acknowledgements}
     Anitha Ravishankar and Grzegorz Micha$\l$ek were supported by NCN through the grant UMO-2017/25/B/ST9/00536 and DSC grant N17/MNS/000038. The authors thank the referee for the useful comments and suggestions that have greatly improved the quality of the manuscript. We thank all the members of the SOHO/LASCO, STEREO/SECCHI and GOES consortium who built the instruments and provided the data used in this study.
     
\end{acknowledgements}

%

\begin{thebibliography}{}

\bibitem[Berkebile et al., (2012)] {berkebile12} Berkebile-Stoiser, S., Veronig, A.~M., Bein, B.~M., et al.\ 2012, \apj, 753, 88.

\bibitem[Bronarska and Micha$\l$ek (2018)] {bronarska}  Bronarska, K., \& Michalek, G.\ 2018, Advances in Space Research, 62, 408.

\bibitem[Bougeret et al., (1995)]{bougeret} Bougeret, J.-L., Kaiser, M.~L., Kellogg, P.~J., et al.\ 1995, \ssr, 71, 231.

\bibitem[Brueckner et al.(1995)]{Brueckner95} Brueckner, G.~E., Howard, R.~A., Koomen, M.~J., et al.\ 1995, \solphys, 162, 357.

\bibitem[Cane et al., (1986)] {cane1986} Cane, H.~V., McGuire, R.~E., \& von Rosenvinge, T.~T.\ 1986, \apj, 301, 448.

\bibitem[Cane et al., (1987)] {cane1987} Cane, H.~V., Sheeley, N.~R., \& Howard, R.~A.\ 1987, \jgr, 92, 9869.

\bibitem[Cane et al., (2000)] {cane2000} Cane, H.~V., Richardson, I.~G., \& St. Cyr, O.~C.\ 2000, \grl, 27, 3591.

\bibitem[Cliver et al., (1999)] {cliver} Cliver, E.~W., Webb, D.~F., \& Howard, R.~A.\ 1999, \solphys, 187, 89.

\bibitem[Dulk and McLean (1978)] {dulk}  Dulk, G.~A., \& McLean, D.~J.\ 1978, \solphys, 57, 279.

\bibitem[Eselevich and Eselevich (2008)] {eselevich} Eselevich, M.~V., \& Eselevich, V.~G.\ 2008, \grl, 35, L22105.

\bibitem[Gleisner and Watermann(2006)]{gleisner} Gleisner, H., \& Watermann, J.\ 2006, Space Weather, 4, S06006.

\bibitem[Gopalswamy et al., (2001a)] {gopalswamy2001a} Gopalswamy, N., Yashiro, S., Kaiser, M.~L., et al.\ 2001, \apjl, 548, L91.


\bibitem[Gopalswamy et al., (2001b)] {gopalswamy2001b} Gopalswamy, N., Lara, A., Yashiro, S., et al.\ 2001, \jgr, 106, 29207.

\bibitem[Gopalswamy et al., (2001c)] {gopalswamy2001c} Gopalswamy, N., Lara, A., Kaiser, M.~L., et al.\ 2001, \jgr, 106, 25261.

\bibitem[(2002a)] {2002a} Gopalswamy, N., Yashiro, S., Micha{\l}ek, G., et al.\ 2002, \apjl, 572, L103.

\bibitem[Gopalswamy (2002)] {gopalswamy2002} Gopalswamy, N.\ 2002, Solar-terrestrial Magnetic Activity and Space Environment, 157.

\bibitem[ (2002b)] {gopalswamy2002b} Gopalswamy, N., Yashiro, S., Kaiser, M.~L., et al.\ 2002, \grl, 29, 1265.

\bibitem[Gopalswamy et al., (2003)] {gopalswamy2003} Gopalswamy, N., Yashiro, S., Lara, A., et al.\ 2003, \grl, 30, 8015.

\bibitem[Gopalswamy et al., (2004)] {gopalswamy2004} Gopalswamy, N., Yashiro, S., Krucker, S., et al.\ 2004, Journal of Geophysical Research (Space Physics), 109, A12105.


\bibitem[Gopalswamy et al., (2009)] {gopalswamy2009} Gopalswamy, N., Yashiro, S., Michalek, G., et al.\ 2009, Earth Moon and Planets, 104, 295.

\bibitem[Gopalswamy (2012)] {gopalswamy2012} Gopalswamy, N.\ 2012, American Institute of Physics Conference Series, 247.

\bibitem[Gopalswamy et al., (2015)]{gopalswamy2015} Gopalswamy, N., M{\"a}kel{\"a}, P., Yashiro, S., et al.\ 2015, Journal of Physics Conference Series, 012012.

\bibitem[Gosling (1991)] {gosling1991} Gosling, J.~T., McComas, D.~J., Phillips, J.~L., et al.\ 1991, \jgr, 96, 7831.

\bibitem[Gosling (1993)] {gosling1993} Gosling, J.~T.\ 1993, \jgr, 98, 18937.

\bibitem[Howard et al.(2008)]{Howard08} Howard, R.~A., Moses, J.~D., Vourlidas, A., et al.\ 2008, \ssr, 136, 67.

\bibitem[Kahler (2001)] {kahler2001} Kahler, S.~W.\ 2001, \jgr, 106, 20947.

\bibitem[Kahler and Vourlidas (2005)] {kahler2005} Kahler, S.~W., \& Vourlidas, A.\ 2005, Journal of Geophysical Research (Space Physics), 110, A12S01.

\bibitem[Kahler \& Vourlidas(2013)]{kahler13} Kahler, S.~W., \& Vourlidas, A.\ 2013, \apj, 769, 143.

\bibitem[Kim et al., (2005)] {kim} Kim, R.-S., Cho, K.-S., Moon, Y.-J., et al.\ 2005, Journal of Geophysical Research (Space Physics), 110, A11104.

\bibitem[LeBlanc et al.,(1998)] {leblanc} Leblanc, Y., Dulk, G.~A., \& Bougeret, J.-L.\ 1998, \solphys, 183, 165.

\bibitem[Li et al.(2012a)]{Li2012a} Li, G., Ao, X., Verkhoglyadova, O., et al.\ 2012, American Institute of Physics Conference Series, 178.

\bibitem[Li et al.(2012b)]{Li2012b} Li, G., Zank, G., Verkhoglyadova, O., et al.\ 2012, American Institute of Physics Conference Series, 115.

\bibitem[Mann et al.,(1999)] {mann} Mann, G., Aurass, H., Klassen, A., et al.\ 1999, 8th SOHO Workshop: Plasma Dynamics and Diagnostics in the Solar Transition Region and Corona, 477.

\bibitem[Manoharan et al., (2004)] {manoharan2004} Manoharan, P.~K., Gopalswamy, N., Yashiro, S., et al.\ 2004, Journal of Geophysical Research (Space Physics), 109, A06109.

\bibitem[Manoharan (2006)] {manoharan2006} Manoharan, P.~K.\ 2006, \solphys, 235, 345.

\bibitem[(2010)] {2010} Manoharan, P.~K.\ 2010, Solar and Stellar Variability: Impact on Earth and Planets, 356.

\bibitem[Manoharan and Mujiber Rahman (2011)] {manoharan2011} Manoharan, P.~K., \& Mujiber Rahman, A.\ 2011, Journal of Atmospheric and Solar-Terrestrial Physics, 73, 671.


\bibitem[Miteva et al., (2013)] {miteva} Miteva, R., Klein, K.-L., Malandraki, O., et al.\ 2013, \solphys, 282, 579.

\bibitem[Michalek et al.(2017)]{Michalek17} Michalek, G., Gopalswamy, N., \& Yashiro, S.\ 2017, \solphys, 292, 114.

\bibitem[Moon et al., (2005)] {moon} Moon, Y., Cho, K., Dryer, M., et al.\ 2005, AGU Spring Meeting Abstracts 2005, SH23A-01.

\bibitem[Nipa et al., (2013)] {nipa} Bhatt, N.~J., Jain, R., \& Awasthi, A.~K.\ 2013, Research in Astronomy and Astrophysics, 13, 978-990.

\bibitem[Pande et al., (2018)] {Pande} Pande, B., Pande, S., Chandra, R., et al.\ 2018, Advances in Space Research, 61, 777.

\bibitem[Ravishankar and Micha$\l$ek (2019)] {ravishankar} Ravishankar, A., \& Micha{\l}ek, G.\ 2019, \solphys, 294, 125.

\bibitem[Reames (1999)] {reames1999} Reames, D.~V.\ 1999, \ssr, 90, 413.

\bibitem[Reames (2013)] {reames2013} Reames, D.~V.\ 2013, \ssr, 175, 53.

\bibitem[Richardson et al.(2014)]{Richardson14} Richardson, I.~G., von Rosenvinge, T.~T., Cane, H.~V., et al.\ 2014, \solphys, 289, 3059.

\bibitem[Richardson et al.(2015)]{Richardson15} Richardson, I.~G., von Rosenvinge, T.~T., \& Cane, H.~V.\ 2015, \solphys, 290, 1741.

\bibitem[Shanmugaraju et al., (2015)] {shanmugaraju} Shanmugaraju, A., Syed Ibrahim, M., Moon, Y.-J., et al.\ 2015, \solphys, 290, 1417.


\bibitem[Sheeley et al., (1997)] {sheeley} Sheeley, N.~R., Wang, Y.-M., Hawley, S.~H., et al.\ 1997, \apj, 484, 472.

\bibitem[Srivastava and Venkatakrishnan (2002)] {srivastava} Srivastava, N., \& Venkatakrishnan, P.\ 2002, \grl, 29, 1287.


\bibitem[St. Cyr, O. C et al., (2000)] {St} St. Cyr, O.~C., Plunkett, S.~P., Michels, D.~J., et al.\ 2000, \jgr, 105, 18169.

\bibitem[Temmer et al., (2008)] {temmer} Temmer, M., Veronig, A.~M., Vr{\v{s}}nak, B., et al.\ 2008, \apjl, 673, L95.

\bibitem[Vourlidas et al.(2010)]{Vourlidas10} Vourlidas, A., Howard, R.~A., Esfandiari, E., et al.\ 2010, \apj, 722, 1522.

\bibitem[Watanabe et al., (2012)] {watanabe} Watanabe, K., Masuda, S., \& Segawa, T.\ 2012, \solphys, 279, 317.

\bibitem[Webb et al., (2000)]{webb} Webb, D.~F., Cliver, E.~W., Crooker, N.~U., et al.\ 2000, \jgr, 105, 7491.

\bibitem[Xie et al.(2019)]{Xie19} Xie, H., St. Cyr, O.~C., M{\"a}kel{\"a}, P., et al.\ 2019, Journal of Geophysical Research (Space Physics), 124, 6384.

\bibitem[Yashiro et al., (2004)] {yashiro2004} Yashiro, S., Gopalswamy, N., Michalek, G., et al.\ 2004, Journal of Geophysical Research (Space Physics), 109, A07105.

\end{thebibliography}
%

\end{document}